\documentclass{osa-article}

\journal{oe}

\usepackage{lineno}
\usepackage{gensymb} 

\begin{document}

\title{Intrinsically accurate sensing with an optomechanical accelerometer}

\author{Benjamin J. Reschovsky\authormark{1,3}, David A. Long\authormark{1,4}, Feng Zhou\authormark{1,2}, Yiliang Bao\authormark{1,2}, Richard A. Allen\authormark{1}, Thomas W. LeBrun\authormark{1,5}, and Jason J. Gorman\authormark{1,6}}

\address{\authormark{1}National Institute of Standards and Technology, 100 Bureau Drive, Gaithersburg, Maryland 20899, USA\\
\authormark{2}Theiss Research, La Jolla California 92037, USA\\
\authormark{3}ben.reschovsky@nist.gov\\
\authormark{4}david.long@nist.gov\\
\authormark{5}lebrun@nist.gov\\
\authormark{6}gorman@nist.gov}

\begin{abstract*}
We demonstrate a microfabricated optomechanical accelerometer that is capable of percent-level accuracy without external calibration. To achieve this capability, we use a mechanical model of the device behavior that can be characterized by the thermal noise response along with an optical frequency comb readout method that enables high sensitivity, high bandwidth, high dynamic range, and SI-traceable displacement measurements. The resulting intrinsic accuracy was evaluated over a wide frequency range by comparing to a primary vibration calibration system and local gravity. The average agreement was found to be 2.1~\% for the calibration system between 0.1~kHz and 15~kHz and better than 0.2~\% for the static acceleration. This capability has the potential to replace costly external calibrations and improve the accuracy of inertial guidance systems and remotely deployed accelerometers. Due to the fundamental nature of the intrinsic accuracy approach, it could be extended to other optomechanical transducers, including force and pressure sensors.
\end{abstract*}

\section{Introduction}
Optomechanical sensors have been widely explored for precision measurements of force \cite{Gavartin2012, Melcher2014}, mass\cite{Sansa2020, Djorwe2019}, and ultrasound \cite{Basiri-Esfahani2019, Westerveld2021}, among others, due to their exceptional displacement sensitivity. Similarly, recent advances have demonstrated the benefits of optomechanical accelerometers, including excellent sensitivity ($< 10~\mu g_n/\mathrm{Hz}^{1/2}$, where $g_n =9.80665$ m/s$^2$ is the standard value of the acceleration due to gravity)\cite{Krause2012, GuzmanCervantes2014, Gerberding2015, Li2018, Huang2020, Zhou2021}, immunity to electromagnetic noise, wide bandwidth operation\cite{Krause2012, GuzmanCervantes2014, Gerberding2015, Zhou2021, Li2018a}, low-frequency operation \cite{Krause2012, Huang2020, Zhou2021}, and scalable manufacturing using microfabrication \cite{Krause2012,Huang2020,Zhou2021}. In addition, by incorporating an optical interferometer in the device, this technology has the potential to perform accurate and SI-traceable measurements without an external calibration \cite{GuzmanCervantes2014, Gerberding2015, Zhou2021, Pratt2021}. We refer to devices with this capability as intrinsically accurate. In contrast, traditional electromechanical devices have voltage or current readouts, requiring external calibration of the sensor output to determine acceleration.  

An external calibration uses an acceleration reference provided by gravity  \cite{ISO16063-16}, a shaker platform in conjunction with a laser interferometer \cite{ISO16063-11}, a centrifuge \cite{ISO5347-8}, a reciprocity procedure \cite{ISO16063-12}, or a secondary reference accelerometer \cite{ISO16063-21}. Although the achievable uncertainty depends on the quality and frequency of calibrations as well as the stability of the accelerometer, National Metrological Institutes (NMIs) typically achieve a $1 \sigma$ relative uncertainty between 0.05~\% and 1.5~\% for vibrations between 100~Hz and 20~kHz\cite{Bruns2021}. However, most deployed accelerometers are calibrated using secondary methods at  higher uncertainty levels\cite{Dytran2016, Keysight2018, ModalShop2020}. These external calibrations impose additional costs, necessitate system downtime, and are prone to errors due to sensitivity drift between calibrations.

By incorporating an optical microcavity into the sensor, the optomechanical accelerometer directly measures the displacement of a proof mass using the precise and SI-traceable laser wavelength as a length reference. Then, the applied acceleration is calculated using a model of the proof mass mechanical response. Ideally, the sensor is designed such that a simple single-mode harmonic oscillator model provides sufficient accuracy, in which case the model is defined by two parameters: the resonance frequency, $\omega_0$, and mechanical quality factor, $Q$, of the accelerometer’s fundamental mechanical mode. These parameters can be characterized by an additional optical measurement, which can be performed in situ with little additional equipment \cite{GuzmanCervantes2014, Gerberding2015, Zhou2021, Pratt2021}. The convenience of this mechanical model characterization step means that it can be performed frequently, potentially improving accuracy performance over current methods by reducing errors due to drift between calibrations. This technique could enable unprecedented accuracy for remotely deployed sensors that cannot be recalibrated and for important applications such as gravimetry and inertial navigation, where maintaining a fixed uncertainty throughout the lifetime of the sensor is critical. 
One prior study has evaluated the intrinsic accuracy of an optomechanical accelerometer using a primary reference \cite{Pratt2021}. That test focused on low frequency vibrations from 3~Hz to 30~Hz and achieved an agreement with an external laser interferometer reference to within approximately 1~\%. In addition, two prior works have also discussed the intrinsic accuracy of optomechanical accelerometers \cite{GuzmanCervantes2014, Gerberding2015}, all using a similar device design with an optical resonator combined with a high-$Q$ mechanical resonator using a parallelogram suspension. Notably, these previous implementations have all employed a side-fringe detection scheme that uses the slope of an optical resonance to transduce the frequency shifts associated with proof mass displacements into laser intensity fluctuations. In addition to limiting the dynamic range of the sensor, this technique requires \textit{a priori} knowledge of the optical lineshape slope. Therefore, an additional calibration step is required to convert the measured photodiode voltage to the units of acceleration. 

We note that the previous work on this topic has used different terminology, such as ‘self-calibrated’ \cite{Gerberding2015} and ‘in-situ calibrated' \cite{Pratt2021} to refer to similar approaches. However, we prefer the term ‘intrinsic accuracy’ to emphasize that the technique involves performing a primary acceleration measurement instead of a comparison to an acceleration reference as implied by the term ‘calibration.’ 

In this paper, we evaluate the intrinsic accuracy of a novel optomechanical accelerometer with unique optical and mechanical properties for a broad range of test frequencies from 100~Hz to 15~kHz, as well as for static accelerations. Our device features a highly-symmetric, microfabricated, low-$Q$ mechanical resonator coupled with a high-finesse optical microcavity \cite{Zhou2021}. This regime allows us to use thermomechanical noise to characterize the mechanical model, which avoids the necessity of an external excitation, as used in previous demonstrations \cite{GuzmanCervantes2014, Pratt2021}. We also present the first implementation of  an  optical frequency comb-based readout method \cite{Long2021} to determine the proof mass displacement using only optical and RF frequencies as references. This approach directly measures acceleration in the SI units of distance and time and avoids the dynamic range and accuracy limitations of the side-fringe method. Combining the accelerometer design with the frequency comb readout method results in the widest bandwidth optomechanical accelerometer with percent-level accuracy, the first to achieve accuracies near a part in ${10}^3$ for the case of static accelerations, and the first to offer direct SI-traceability. 

In the following section, we describe the details of the optomechanical accelerometer and displacement readout techniques. We then present the details of the intrinsic accuracy approach in Section \ref{sec:intrinsic}. Section \ref{sec:results} provides our accuracy evaluation results for two different acceleration references. In Section \ref{sec:shaker} we use the National Institute of Standards and Technology (NIST) primary vibration calibration system (PVCS) to examine the optomechanical accelerometer’s intrinsic accuracy for broadband sinusoidal vibrations. In Section \ref{sec:gravity} we use local gravity to test the accuracy for static accelerations. Finally, we discuss our conclusions in Section \ref{sec:conclusions}.

\section{Experimental Methods\label{sec:methods}}
\subsection{Optomechanical accelerometer\label{sec:accel}}

\begin{figure}

\centering\includegraphics[width=\columnwidth]{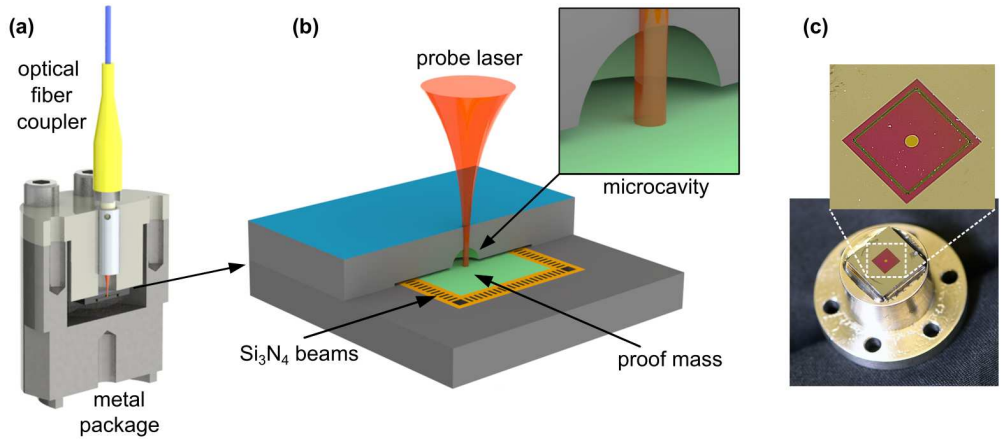}
\caption{\label{fig:accel} Accelerometer design. (a) Cutaway of the complete assembly, including the microfabricated components, metal package, and optical coupler. (b) An enlarged view of the microfabricated silicon chips. The top chip contains a concave micromirror (shown in cross-section) and the bottom chip features a silicon proof mass suspended by Si$_3$N$_4$ microbeams. The interior surfaces have high reflectivity mirror coatings (green), and the exterior surfaces have anti-reflective coatings (blue). (c) Photograph of the silicon chip stack assembled on the mount without the cover. Details such as the anti-reflective coating (yellow), Si$_3$N$_4$ microbeams (green), Si$_3$N$_4$ layer on silicon (red), and exposed silicon surface (gray) are visible.}
\end{figure}

The optomechanical accelerometer used in this study includes a microscale hemispherical Fabry-P\'erot cavity that is formed by two silicon chips (see Fig.~\ref{fig:accel}). The flat mirror of the cavity is incorporated into a millimeter-scale proof mass that is suspended on the top and bottom by silicon nitride (Si$_3$N$_4$) microbeams, forming a mechanical resonator. When an acceleration is applied to the chip assembly, the proof mass displaces, thereby changing the cavity length, which can be measured with an optical cavity readout method. The proof mass is nominally 4~mm $\times$ 4~mm $\times$ 0.525~mm with a mass of approximately 20~mg. The supporting beams are 20~$\mu$m wide, 84~$\mu$m long, and spaced by 20~$\mu$m giving the structure a resonance frequency ($\omega_0/2\pi$) of approximately 7.85~kHz and a $Q$ of 16 in air. The other mirror is a concave silicon micromirror that is etched into a second chip. The two mirrors have high-reflectivity dielectric coatings (alternating Ta$_2$O$_5$ and SiO$_2$ layers) that yield a cavity finesse greater than 3000 and an approximate cavity mode linewidth (full width at half maximum) of $\Gamma = 185$~MHz. The cavity length is approximately 240~µm, resulting in a free spectral range of 620~GHz. Further details on the accelerometer fabrication, design, and performance can be found in Refs. \cite{Bao2017,Zhou2021}.

The two silicon chips are bonded to a stainless-steel package using UV-curable epoxy. The package also houses an optical coupler that couples light from a polarization-maintaining (PM) optical fiber into and out of the cavity. The package includes a metal cover that protects the device and allows for convenient mounting. To keep the total weight of the device relatively low for vibration measurements, the mount cover is fabricated from aluminum, leading to a total mass of 67.0(2)~g. All measurements in this paper are conducted in air at room temperature.

\subsection{Displacement readout\label{sec:readout}}

\begin{figure}

\includegraphics[width=\columnwidth]{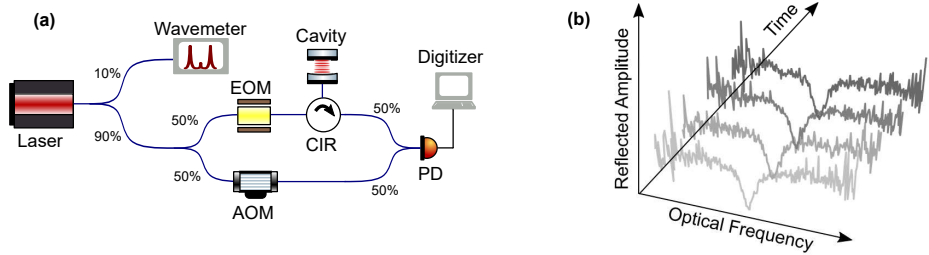}
\caption{\label{fig:comb} (a) Diagram of the electro-optical frequency comb readout method used to measure the optical cavity frequency shifts, where EOM: electro-optic modulator, AOM: acousto-optic modulator, CIR: circulator, and PD: photodetector. (b) The digitized data is divided into sub-interferograms, each comprising 1~$\mu$s of measurement time, which are processed to reveal the cavity mode spectrum relative to the laser frequency. Four examples of these processed spectra are presented, showing how they are used to track the cavity mode position as a function of time.}
\end{figure}

We used two cavity readout methods, one to monitor the displacement of the accelerometer proof mass during acceleration measurements and the other to measure the thermomechanical noise response of the proof mass to characterize the mechanical parameters. The first method uses an electro-optic frequency comb to perform rapid, high-dynamic range optical readout of the cavity displacement (see Fig.~\ref{fig:comb}). This is the first application of an optical frequency comb for evaluating the accuracy of an optomechanical sensor, although the details of the comb generation have been described previously \cite{Long2016, Long2017, Long2019, Long2021}. We tuned an external-cavity diode laser (ECDL) near an optical resonance of the optical cavity at approximately 1547~nm. An electro-optic phase modulator was driven with a repeating train of linear frequency chirps\cite{Long2016, Long2019}, generating an optical frequency comb with a tooth spacing of 10~MHz and a span of 2.2~GHz, centered on the laser frequency. This optical frequency comb was reflected off of the optical cavity and detected on a fast photodetector using a self-heterodyne architecture\cite{Long2016, Hebert2015}. We digitized the photodetector signal with a sampling rate of 3~GS/s. The resulting interferogram, with a duration of 0.5~s, was divided into sub-interferograms of 3000~samples corresponding to a sampling time of 1~$\mu$s. We Fourier transformed each of these sub-interferograms to generate a cavity mode spectrum . These spectra were then fit with an asymmetric Fano lineshape\cite{Fano1961} to extract the cavity mode frequency, relative to the laser carrier frequency, as a function of time (see Fig.~\ref{fig:comb}(b)). 

The measured optical frequency shifts, $\delta\nu$, were converted to cavity displacements using the relation\cite{Lawall2005}
\begin{equation} \label{eq:displacement}
    x=\frac{c}{2n\left(\nu_0+\delta\nu\right)}\left[\frac{1}{\pi}\left(\Phi_{G,1}-\Phi_{G,0}\right)-\frac{\delta\nu}{\Delta\nu_{\mathrm{FSR,0}}}\right],
\end{equation}
	
where $\nu_{\mathrm{FSR,0}}$ is the cavity’s free spectral range and $\nu_0$ is the initial optical frequency of the cavity mode, both of which we measure using a high-accuracy wavemeter. We used the Ciddor equation\cite{Ciddor1996} to calculate the index of refraction, $n$, based on the ambient conditions in the laboratory. We note that Eq.~(\ref{eq:displacement}) also accounts for the difference between the initial and final Gouy phases, $\Phi_{G,0}$ and $\Phi_{G,1}$, which results in a small correction on the order of ${10}^{-4}$\cite{Lawall2005}. See the supplementary material for more details.

\begin{figure}

\includegraphics[width=\columnwidth]{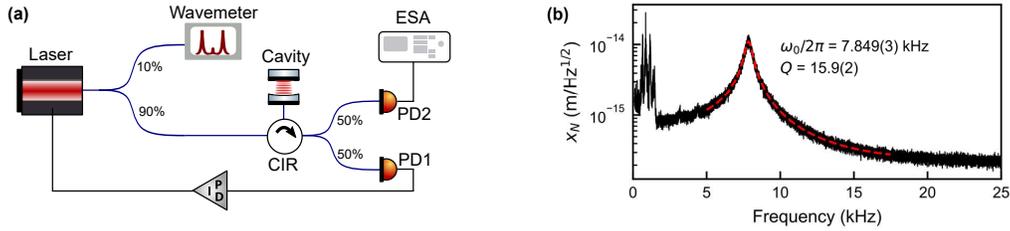}
\caption{\label{fig:sidelock} (a) Diagram of the optical side-fringe method used to measure the displacement noise spectrum of the accelerometer, including the thermomechanical noise, ESA: electronic spectrum analyzer, and PID: proportional-integral-derivative controller. Photodiode~1 (PD1) is used for a side-of-fringe lock using a slow PID controller (integrator cutoff frequency = 100~Hz). Cavity displacements faster than the loop bandwidth are detected as intensity fluctuations by PD2 and the ESA. (b) The displacement power spectral density due to thermomechanical motion. The dashed line is a fit to Eq.~(\ref{eq:noise_fit}), which is used to determine $\omega_0$ and $Q$.}
\end{figure}

The second readout method uses a side-fringe transduction method to measure the proof-mass thermal noise spectrum. To cancel long term drifts of the optical cavity or laser, the laser was locked to the side of the optical resonance at a detuning of approximately $0.3\Gamma$ with a low-bandwidth feedback loop ($\cong 100$~Hz, see Fig.~\ref{fig:sidelock}(a)). Cavity displacements due to the thermal Brownian motion of the proof mass at frequencies faster than the feedback bandwidth are measured with the reflected light intensity. These fluctuations were detected by a photodiode and collected with a spectrum analyzer using an averaging time of 75~s. A measurement of the lineshape slope at the controller lock point is used to convert the photodetector voltage to the displacement noise floor shown in Fig.~\ref{fig:sidelock}(b).

The side-fringe method is commonly used to transduce the proof mass displacements into a photodiode voltage\cite{Krause2012, GuzmanCervantes2014, Gerberding2015, Li2018, Li2018a} and is well suited for the noise spectrum measurement because it can achieve a low noise floor for small amplitude signals. For our apparatus, the side-fringe method has a noise floor of $1.5\times{10}^{-16}$~m/Hz$^{1/2}$ while the comb readout detection limit is $5\times{10}^{-15}$~m/Hz$^{1/2}$. The accelerometer sensitivity is largely limited by thermomechanical noise from the mechanical resonator when using the side-fringe method and by readout noise when using the frequency comb method. However, there are several advantages of the optical frequency comb method for the actual acceleration measurements. For instance, it enables detection of large shifts in the cavity resonance frequency compared to its linewidth, with a full range of 2.2~GHz. Therefore, we can use a narrow linewidth cavity to enhance sensitivity while also detecting large amplitude displacements. Previous work has demonstrated a linear dynamic range of over four orders of magnitude\cite{Long2021} and can likely be extended one to two orders of magnitude further. In addition, the side-fringe method requires \textit{a priori} knowledge of the lineshape slope at the lockpoint, which is difficult to determine at the percent level or better. This does not limit our ability to use the noise spectrum to characterize the mechanical model since that process is insensitive to the overall amplitude of the noise spectrum as will be described in Section \ref{sec:intrinsic}, but it would be a significant error source if we were to use the side-fringe method during acceleration measurements. In contrast, the comb readout method measures the proof mass displacement solely in terms of optical and RF frequencies, both of which can be determined with very high accuracy. For example, we measured the optical frequency using a wavemeter with a relative uncertainty of $<{10}^{-7}$ and the 10~MHz comb tooth spacing was referenced to a Rb atomic clock with an uncertainty of $<{10}^{-9}$. The wavemeter was calibrated with respect to a polarization-stabilized He-Ne laser, which offered a stability better than 100~kHz for a 10~ms timescale. This approach provides a straightforward path to SI-traceability and does not rely on characterizing the slope of a resonance feature as is necessary with the side-fringe method.

\section{Intrinsic accuracy approach\label{sec:intrinsic}}

\subsection{Displacement-to-acceleration conversion}

To perform an accurate acceleration measurement, we must convert the detected proof mass displacement to an acceleration. We therefore define a mechanical model that describes the motion of the accelerometer’s proof mass, which is expressed in Fourier space as $x\left(\omega\right)=G\left(\omega\right)a\left(\omega\right)$, where $a\left(\omega\right)$ is the applied acceleration at angular frequency $\omega$ and $G\left(\omega\right)$ is the mechanical susceptibility\cite{Zhou2021}. This relation can be inverted to calculate the applied acceleration as $a\left(\omega\right)={G\left(\omega\right)}^{-1}x\left(\omega\right)$. For this paper, we measure acceleration amplitudes for static and sinusoidal excitations, which are calculated using
\begin{equation} \label{eq:inversion}
    \left|a\left(\omega\right)\right|=\left|G\left(\omega\right)\right|^{-1}\left|x\left(\omega\right)\right| .	
\end{equation}
The problem of calculating the generic case of an arbitrary time-dependent acceleration can be treated as a deconvolution calculation and requires transforming back into the time domain\cite{Riad1986}.

The proof mass structure in our device has a translational mode (i.e., piston mode) at a resonance frequency of $\omega_0/2\pi \cong 7.85$~kHz, while higher order modes (e.g., rotational and rocking modes) have resonance frequencies above 60~kHz due to the structural design of the mechanical resonator\cite{Zhou2021}. This large separation between the translational and higher-order modes allows us to model the mechanical resonator as a viscously damped, single-mode harmonic oscillator. In this case,
\begin{equation} \label{eq:G}
    G\left(\omega\right)=\frac{1}{\omega_0^2-\omega^2+i\omega\omega_0/Q} ,
\end{equation}
where $\omega_0=2\pi f_0=\left(k/m\right)^{1/2}$, $k$ is the resonator stiffness, and $m$ is the proof mass. 

\subsection{Model characterization\label{sec:model}}

We can fully define the mechanical model, $G\left(\omega\right)$, by measuring two parameters: $\omega_0$ and $Q$. As shown in Fig.~\ref{fig:sidelock}(b) and Ref.~\cite{Zhou2021}, our accelerometer is limited by the thermomechanical noise over a wide frequency range, so we use the noise spectrum to perform this model characterization. First, we measure the displacement spectral density using the side-fringe method described in the Section~\ref{sec:readout} above while the device is excited only by thermal noise. Then, we fit the displacement spectral density using the equation
\begin{equation} \label{eq:noise_fit}
x_N\left(\omega\right)=\sqrt{{x_{th}\left(\omega\right)}^2+x_s\left(\omega\right)^2}=\sqrt{\left|G\left(\omega\right)\right|^2a_{th}^2+s_0^2} ,
\end{equation}
where the equivalent acceleration due to thermomechanical noise is $a_{th}=\left(4k_BT\omega_0/mQ\right)^{1/2}$, $k_B$ is the Boltzmann constant, and $T$ is the temperature\cite{Zhou2021, Gabrielson1993}. The thermal noise term, $x_{th}\left(\omega\right)$, was added in quadrature with the frequency independent shot noise term, $x_s\left(\omega\right)=s_0$, since these noise sources are uncorrelated. The resulting fit function has four fit parameters, $m$, $\omega_0$, $Q$, and $s_0$. Since $m$ and $s_0$ are not needed to define $G\left(\omega\right)$, we can neglect a careful determination of various properties that affect the noise spectrum amplitude such as lineshape slope, laser wavelength, cavity length, temperature, optical power, and photodetector response. 

A typical displacement noise spectrum (black trace) and fit using Eq.~(\ref{eq:noise_fit}) (red dashed line) are shown in Fig.~\ref{fig:sidelock}(b). The piston mode resonance near 7.85~kHz is clearly visible and well-modeled by the fit. The narrow peaks below 2~kHz were due to laser frequency noise originating from resonances in the laser’s external cavity. We note that model characterization based on the thermomechanical noise is possible because the mechanical and optical resonators of our device were designed so that the thermomechanical noise is well above the shot noise detection limit over a wide frequency range\cite{Zhou2021}. Other optomechanical accelerometers\cite{Krause2012, GuzmanCervantes2014, Gerberding2015, Li2018, Huang2020, Li2018a} have been limited by shot noise except for within a narrow frequency range near the mechanical resonance frequency, leading to an increase in the fit uncertainty.

\subsection{Uncertainty analysis\label{sec:uncertainty}}

\begin{figure}

\centering\includegraphics{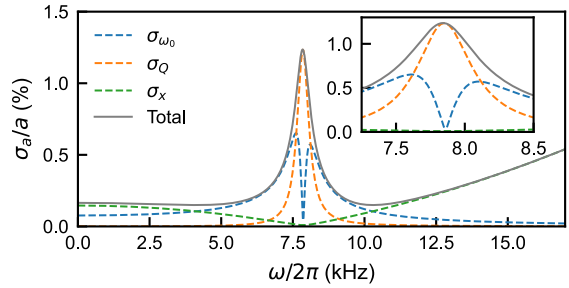}
\caption{\label{fig:uncertainty} The calculated fractional acceleration uncertainty as a function of vibrational frequency for a 1 m/s$^2$ acceleration. The three contributions due to the individual uncertainty in $\omega_0$, $Q$, and $x$, are summed in quadrature. The total uncertainty is less than 0.2~\% below 5~kHz. The uncertainty near resonance is dominated by the uncertainty in $Q$ and above resonance the total uncertainty is driven by the displacement uncertainty as the signal-to-noise ratio degrades.}
\end{figure}

The uncertainty in the measured $x$, $\omega_0$, and $Q$ limit the accuracy to which we can determine an acceleration. We estimated the uncertainty in the measured displacement using a simulated signal added to the measured noise signal, giving a result of $\sigma_x=0.6$~pm for displacement amplitudes greater than 100~pm (see supplementary material). Over the course of 8~hours, we found that the repeatability of $\omega_0$ and $Q$ are $\sigma_{\omega_0}/2\pi=3$~Hz and $\sigma_Q=0.2$, respectively. These standard deviations are approximately one order of magnitude larger than the fit uncertainty for any given thermal noise measurement, so the variation is likely dominated by changing environmental factors. See Fig.~\ref{fig:uncertainty} for the fractional acceleration uncertainty due to these three error sources as well as their quadrature sum for a constant acceleration of 1~m/s$^2$ from DC to 20~kHz (see the supplementary material for more information). Below 5~kHz, we find that $\sigma_a/a < 0.2$~\%. There is a narrow peak in the uncertainty around the resonance frequency due to the contribution of $\sigma_Q$, and the uncertainty gradually increases above resonance as the displacement signal-to-noise ratio degrades. Aside from the narrow peak near resonance, the uncertainty levels shown in Fig.~\ref{fig:uncertainty} are similar or better than those typically achieved by NMIs using accepted primary calibration methods\cite{Bruns2021}.

\section{Results and discussion\label{sec:results}}

\subsection{\label{sec:shaker} Broadband vibration testing}

\begin{figure}
\includegraphics[width=\columnwidth]{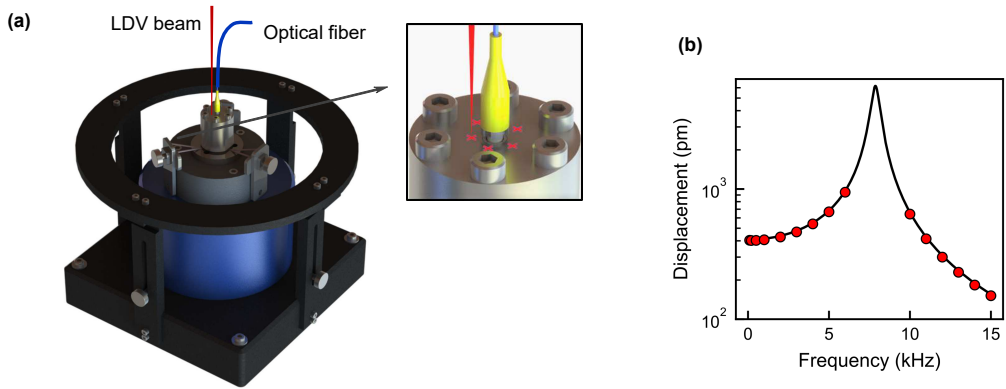}
\caption{\label{fig:shaker1} (a) Experimental setup used for the tests with the primary vibration calibration system (PVCS). We mounted the accelerometer on a high-frequency shaker table and compared the acceleration measured by the optomechanical accelerometer to a laser Doppler vibrometer (LDV) reflected off the top of the accelerometer’s cover. LDV measurements were made at six locations on the accelerometer cover, as indicated by the red marks in the inset. (b) Measured (circles) and calculated (solid line) peak cavity displacement for an applied peak acceleration of 1~m/s$^2$ as a function of vibration frequency. The calculated displacements are given by the inverse of Eq.~(\ref{eq:inversion}) with $\omega_0$ and $Q$ measured from a thermal noise spectrum. The measured uncertainties are smaller than the markers.}
\end{figure}

We evaluated the intrinsic accuracy of the accelerometer for broadband sinusoidal accelerations using the National Institute of Standards and Technology (NIST) primary vibration calibration system (PVCS)\cite{Payne2016}. The system consists of an air-bearing shaker table that provides the sinusoidal motion and a laser Doppler vibrometer (LDV) that provides a low-uncertainty measurement of the acceleration amplitude. See Fig.~\ref{fig:shaker1}(a) for a rendering of the experimental setup. For each set of experimental conditions (acceleration amplitude and frequency), we sampled six different evenly spaced locations on top of the accelerometer with the LDV beam at a radius of 3.75(75)~mm from the center (see Fig.~\ref{fig:shaker1}(a) inset). 

First, we examined the accelerometer performance at a nominal peak acceleration of 1~m/s$^2$ from 0.1~kHz to 6~kHz and from 10~kHz to 15~kHz. These ranges avoid operation near the mechanical resonance at 7.85~kHz to ensure that the proof mass displacement is within the range of the electro-optical frequency comb readout. Recently, we have demonstrated a dual-comb approach that interleaves two frequency combs, increasing the comb span by a factor of ten \cite{Long2022}. This approach can be used in future work to increase the displacement measurement range by an equivalent amount. It is important to note that the accelerometer is capable of operating on resonance, which has been demonstrated previously for smaller accelerations \cite{Zhou2021}. The peak cavity displacements as a function of vibration frequency are shown in Fig.~\ref{fig:shaker1}(b). These measured displacements (red points) agree well both above and below the cavity resonance at 7.85~kHz with the expected displacement from the inverse of Eq.~(\ref{eq:inversion}) with $\left|a\left(\omega\right)\right| = 1$ m/s$^2$ (solid black line). To rule out any contribution of laser noise below 2~kHz (such as seen in Fig.~\ref{fig:sidelock}(b)), we verified that equivalent results are obtained using a low phase-noise fiber laser. 

\begin{figure}

\includegraphics[width=\columnwidth]{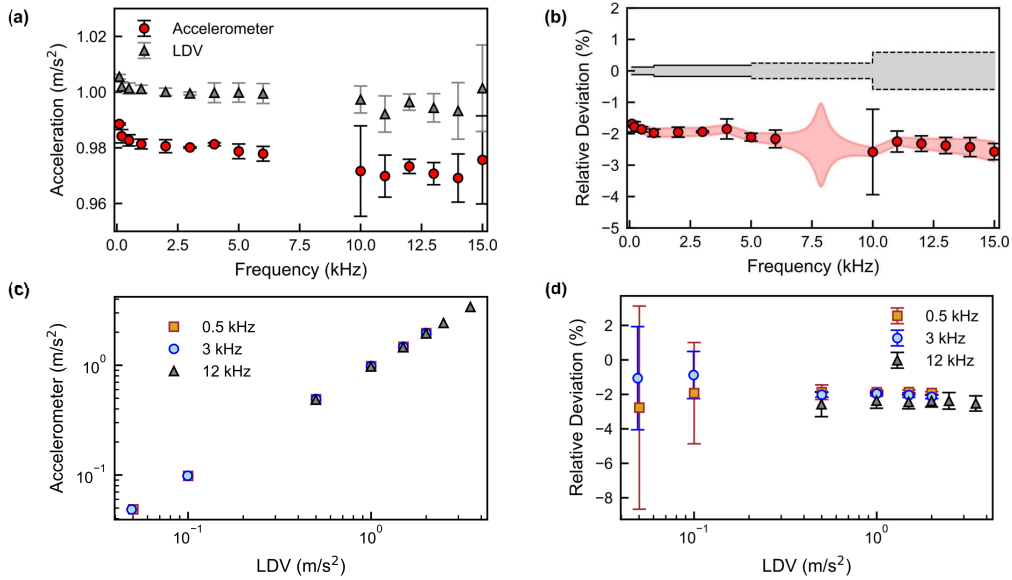}
\caption{\label{fig:shakerData} (a) Acceleration measured by the LDV (gray triangles) and accelerometer (red circles). The error bars indicate the standard deviation from the six different LDV locations. (b) Same data as (a) expressed as a relative deviation between the accelerometer and LDV. The shaded regions indicate the $1\sigma$ uncertainty of the PVCS system (gray) and the uncertainty of the accelerometer (pink). (c) Acceleration measured by the accelerometer (vertical axis) and LDV (horizontal axis) at three different frequencies. The uncertainties are smaller than the markers. (d) Same data as (c) expressed as a relative deviation between the accelerometer and LDV measurements. }
\end{figure}

Next, we evaluated the agreement between the optomechanical accelerometer and the LDV. The accelerations measured by the LDV and accelerometer are shown in absolute units in Fig.~\ref{fig:shakerData}(a) and as a relative percentage in Fig.~\ref{fig:shakerData}(b). The accelerometer results agree well with the primary measurement within 1.7~\% to 2.7~\% from 0.1~kHz to 6~kHz and from 10~kHz to 15~kHz, with the difference increasing slightly at higher frequencies. The average relative agreement across the entire dataset is 2.1(5)~\%. This relative difference is nearly independent of frequency and is slightly larger than the expected uncertainties of the PVCS\cite{Bruns2021} (gray region) and optomechanical accelerometer (pink region) shown in Fig.~\ref{fig:shakerData}(b). The standard PVCS uncertainty budget has been modified slightly to remove inapplicable electronic noise sources used for traditional electromechanical sensors. We note that our measurement conditions fall outside the system’s typical operating regime for frequencies above 5~kHz (shown by the dashed black lines in Fig.~\ref{fig:shakerData}(b)) where our accelerometer is slightly heavier (67~g) than the manufacturer’s maximum specified mass (50~g)\cite{SPEKTRA2016b}.

The accelerometer linearity was also studied at three different vibration frequencies (0.5~kHz, 3~kHz, and 12~kHz) over two orders of magnitude of acceleration from 0.05~m/s$^2$ to 3.5~m/s$^2$ (see Fig.~\ref{fig:shakerData}(c-d)). The results are highly linear with a standard deviation of the residuals of less than 0.0015~m/s$^2$ and a maximum deviation of less than 0.7~\% of the full measurement range. The average relative agreement for the linearity results is 2.1(13)~\%, which is consistent across the amplitudes and frequencies studied. This dataset has a larger standard deviation since the two lowest amplitude points approach the noise floor of the LDV.

We have demonstrated similar performance both significantly above and below the mechanical resonance frequency of our device. This is notable because accelerometers are typically restricted to operating well below their mechanical resonances to ensure that their response is frequency independent. Here, we show that for our device design, a simple mechanical model is sufficient to extend the operating bandwidth up to and beyond the resonance frequency.

As with all primary accelerometer comparisons\cite{Ripper2009, Taubner2010, Bruns2012, Nozato2019, Sprecher2020}, there are many parasitic dynamic effects that can lead to systematic errors. For instance, we found that it was important to consider non-rigid body motion of the accelerometer package. As the vibration frequency increases, the finite stiffness of the materials leads to deformation. As a result, we found that the vibration amplitude was smaller on the top surface of the accelerometer than on the shaker platform and that the accelerometer lid undergoes a drumhead-type motion with larger displacements near the center of the device compared to the exterior of the device. To mitigate these effects, we positioned the LDV beam on top of and as close as possible to the center of the device (as shown in Fig.~\ref{fig:shaker1}(a) inset). Finite-element simulations show that, up to 15~kHz, the relative motion at the locations sampled by the LDV is within 1~\% of the motion on the interior surface to which the silicon chips are attached. Below 6~kHz the differences are $< 0.2$~\%. 

We attribute the increased noise at the 10~kHz vibration frequency to another parasitic dynamic effect: rocking (see Fig.~\ref{fig:shakerData}(b)). At this frequency, one side of the device experiences larger amplitude motion than the other, which is consistent with rocking\cite{Cabral2008}. Finite element simulations of the accelerometer package confirm that a non-symmetric rocking mode can be excited at vibration frequencies near 10~kHz. 

Despite these challenges, we were able to achieve an agreement of approximately 2~\% between the primary LDV measurements and an intrinsically accurate optomechanical accelerometer over a wide range of frequencies and amplitudes. Although this deviation is small, it is several times larger than the expected uncertainty of both the accelerometer and PVCS system as shown in Fig.~\ref{fig:shakerData}(b), suggesting that our performance is limited by an unidentified systematic effect. We believe the most likely cause is the influence of additional mechanical modes, either in the shaker platform or accelerometer packaging that are not captured by our single-mode model. Future work will address the cause and resolution of this systematic effect

\subsection{\label{sec:gravity} Static acceleration testing}

\begin{figure}

\includegraphics[width=\columnwidth]{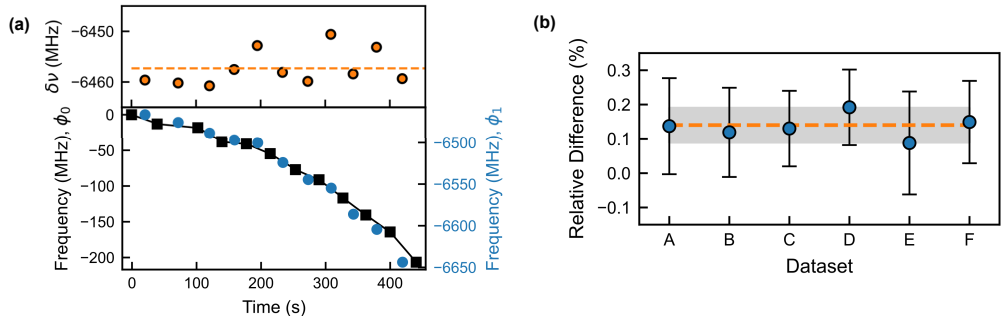}
\caption{\label{fig:gravity} (a) A single dataset in which the accelerometer was repeatedly flipped from an orientation angle of $\phi_0$ (black squares, left vertical axis) to $\phi_1=\phi_0+180 \degree$ (blue circles, right vertical axis), relative to the direction of gravity, while tracking the relative frequency shift of the cavity optical mode. The solid black line shows a linear interpolation of the $\phi_0$ points. The calculated frequency shifts between the blue points and black line are shown in the top panel. The average shift for the dataset (dashed orange line) is $-6457.3$~MHz (b) The relative difference between the gravitational acceleration measured by the optomechanical accelerometer and the reference value for six different datasets taken over the course of two days. The error bars show the combined standard deviation from all non-negligible error sources within each dataset. The weighted average is 0.14~\% (dashed line) with a standard error of 0.05~\% (shaded region). }
\end{figure}

We also evaluated the static performance of the optomechanical accelerometer by using it to measure the local acceleration due to gravity. The accelerometer was placed in a fabricated aluminum enclosure, referred to here as the angle block, that allowed us to rotate the accelerometer axis from being aligned (defined as $\phi=0 \degree$) to anti-aligned ($\phi= 180 \degree$) with respect to gravity. 

Our measurement protocol was as follows: The accelerometer was mounted inside the angle block and oriented at an initial alignment of $\phi=\phi_0$ on top of a large granite table, where $\phi$ is the angle between the accelerometer axis and the direction of gravity and $\phi_0$ is either $0 \degree$ or $180 \degree$. First, we recorded the absolute laser frequency using the wavemeter while simultaneously measuring the relative frequency difference between the laser and cavity resonance using the optical comb readout method (see Fig.~\ref{fig:comb}(a)). Next, we tuned the laser frequency by approximately 6~GHz while also rotating the angle block to $\phi= \phi_1=\phi_0+180\degree$ and repeated the wavemeter and optical comb measurement. Finally, we reversed the process to return to the original laser wavelength and cavity orientation. This process was repeated until we built up a dataset consisting of 21 to 23 total measurements. To allow time for the manual block rotation, each reading is spaced by about 15~s to 20~s. We collected six of these datasets on two different days, with a separate mechanical model characterization for each day. Since, $G\left(0\right)=\omega_0^{-2}$, the quality factor does not contribute to this static measurement.

The results from one of these datasets are shown in Fig.~\ref{fig:gravity}(a). Although we used insulated gloves when handling the aluminum block, the flipping procedure slowly increases the temperature of the cavity due to thermal gradients between the granite table and the ambient air, leading to a slow drift during data capture. This drift, which is approximately quadratic as a function of time, is not observed when the block is at rest. To compensate for the drift, we linearly interpolate between each successive measurement at the $\phi_0$ orientation (illustrated by the black line in Fig.~\ref{fig:gravity}(a)) and calculate the frequency shift relative to this interpolation for every measurement at $\phi_1$ (see the top panel in Fig.~\ref{fig:gravity}(a)). Next, we use Eqs.~(\ref{eq:displacement}) and (\ref{eq:inversion}) to calculate the cavity displacement and acceleration. Finally, we compare the measured acceleration to the known value of $g=9.801018\left(5\right)$~m/s$^2$ based on a detailed measurement previously done at a nearby reference location on the NIST Gaithersburg, Maryland campus\cite{Tate1968}. The correction due to the elevation difference, $h$, between our apparatus and the reference location of $h\left(0.3\times{10}^{-6}\right)$~m/s$^2$ is negligible for this work. All six datasets agree well with each other and the weighted mean differs from the known value of $g$ by only 0.14(5)~\% (see Fig.~\ref{fig:gravity}(b)). This agreement is at least an order of magnitude better than previous tests\cite{GuzmanCervantes2014, Pratt2021} and it demonstrates that instrinsically accurate optomechanical accelerometers have the potential to achieve the sub-percent level uncertainty requirements of demanding applications such as inertial navigation. 

The low uncertainty achieved during static acceleration measurement provides additional evidence that the larger uncertainty found during vibration measurements (Sec.~\ref{sec:shaker}) is due to dynamic effects in the packaging, rather than the cavity readout or model parameters. To the contrary, the largest contributions to the uncertainty of static measurements are the repeatability of the frequency shifts within each dataset and the uncertainty in determining $\omega_0$. These effects could be reduced by straightforward improvements to our measurement routine, such as using a mechanized rotation table to adjust the accelerometer orientation and taking more frequent thermal noise measurements. See the supplementary material for more details on the various sources of error.

We also performed a similar test to put a limit on the cross-axis sensitivity of the accelerometer. In this case, we measured the relative frequency shifts from three different orientation angles, $\phi=0 \degree,\ 90 \degree$, and 180\degree. These tests show that the cross-axis sensitivity for this accelerometer design is quite small at $< 0.4$~\%, potentially limited by small angular errors in the angle block and/or accelerometer packaging.

\section{Conclusions\label{sec:conclusions}}

This paper presents the first evaluation of an intrinsically accurate optomechanical accelerometer for broadband vibrations and the first comparison to an NMI-grade accelerometer calibration system. Also, unlike prior demonstrations\cite{GuzmanCervantes2014,Pratt2021}, this evaluation features a highly symmetric mechanical oscillator embedded in a microfabricated high finesse optical cavity and used an optical frequency comb readout approach. Our device achieved an average agreement of 2.1(5)~\% for vibrations from 0.1~kHz to 15~kHz and for varying amplitude over two orders of magnitude. This performance was achieved over a much wider frequency range than have been previously examined. We note that Ref.~\cite{Pratt2021} also observed a systematic offset compared to the reference on the percent level. The devices in that study feature a very different optical and mechanical design compared to our accelerometer, although both approaches make the same single mode approximation. This similar behavior in an otherwise very different system suggests that additional mechanical modes may be limiting the performance. 

We also present the first evaluation of the intrinsic accuracy approach for static accelerations. In those tests, we reached an agreement of 0.14~\%, which is at least an order of magnitude more accurate than all previous demonstrations. The improved performance compared to our vibration results suggest that any parasitic mechanical modes are only excited by the shaker motion and not while the cavity is at rest.

We attribute this performance to several key features of our accelerometer design: the simple mechanical response, a noise floor that is limited by thermomechanical noise over a large range, and the optical frequency comb readout method. We expect that the sinusoidal vibration accuracy can be improved further by stiffer packaging to reduce the effect of higher order mechanical modes and that the agreement with gravity can be increased by improving the thermal stability during the tests. 

These results demonstrate the potential for intrinsically accurate optomechanical accelerometers to reduce operating costs by eliminating the need for external calibrations while also improving performance. These benefits would be especially valuable for remotely deployed sensors and for applications that demand high accuracy such as inertial navigation and gravimetry. We note that by characterizing additional parameters (i.e., the proof mass and sensor area) this approach could be extended to other types of optomechanical devices, such as force and pressure sensors. Therefore, optomechanics enables a new approach to accurate sensing based on the intrinsic properties of the sensor instead of comparisons to external excitations, expanding these benefits to a myriad of applications.

\begin{backmatter}

\bmsection{Acknowledgments}
This work was partially supported by the NIST on a Chip Program. Y. B. acknowledges support from the National Institute of Standards and Technology (NIST), Department of Commerce, USA (70NANB17H247, 70NANB20H149). F. Z. acknowledges support from the National Institute of Standards and Technology (NIST), Department of Commerce, USA (70NANB20H174). The authors would like to thank Akobuije Chijioke, John Lawall, Ramgopal Madugani, and Jon Pratt for helpful discussions as well as Glenn Holland for the design of the angle block. This research was performed in part in the NIST Center for Nanoscale Science and Technology Nanofab.

\bmsection{Disclosures}
The authors declare no conflicts of interest.

\bmsection{Data availability} Data underlying the results presented in this paper may be obtained from the authors upon reasonable request.

\bmsection{Supplemental document}
See Supplement 1 for supporting content. 

\end{backmatter}


\bibliography{PrimaryPaper,PrimaryPaper2}
\end{document}